\documentclass[sigconf]{acmart}
\usepackage{algorithm}
\usepackage{algorithmic}
\usepackage{adjustbox}
\AtBeginDocument{%
  \providecommand\BibTeX{{%
    \normalfont B\kern-0.5em{\scshape i\kern-0.25em b}\kern-0.8em\TeX}}}
\setcopyright{acmcopyright}
\copyrightyear{2022}
\acmYear{2022}
\acmDOI{}

\acmConference[]{ICAIF'22}{Small Data, Big Opportunities: Making the Most of AI}{2022}
\acmBooktitle{ICAIF, Small Data, Big Opportunities: Making the Most of AI}
\acmPrice{}
\acmISBN{}

\begin{document}

\title{InProC: Industry and Product/Service Code Classification}


\author{Simerjot Kaur}
\affiliation{%
  \institution{JPMorgan AI Research}
    \city{Palo Alto}
  \state{California}
  \country{USA}}
\email{simerjot.kaur@jpmchase.com}

\author{Andrea Stefanucci}
\affiliation{%
  \institution{JPMorgan AI Research}\city{New York}
  \state{NY}
  \country{USA}}
\email{andrea.stefanucci@jpmorgan.com}

\author{Sameena Shah}
\affiliation{%
  \institution{JPMorgan AI Research}\city{New York}
  \state{NY}
  \country{USA}}
\email{sameena.shah@jpmchase.com}

\renewcommand{\shortauthors}{Kaur, et al.}

\begin{abstract}
  Determining industry and product/service codes for a company is an important real-world task and is typically very expensive as it involves manual curation of data about the companies. Building an AI agent that can predict these codes automatically can significantly help reduce costs, and eliminate human biases and errors. However, unavailability of labeled datasets as well as the need for high precision results within the financial domain makes this a challenging problem. In this work, we propose a hierarchical multi-class industry code classifier with a targeted multi-label product/service code classifier leveraging advances in unsupervised representation learning techniques. We demonstrate how a high quality industry and product/service code classification system can be built using extremely limited labeled dataset. We evaluate our approach on a dataset of more than 20,000 companies and achieved a classification accuracy of more than 92\%. Additionally, we also compared our approach with a dataset of 350 manually labeled product/service codes provided by Subject Matter Experts (SMEs) and obtained an accuracy of more than 96\% resulting in real-life adoption within the financial domain. 
\end{abstract}

%



\maketitle

\section{Introduction}\label{intro}
Determining the industry sectors to which a company belongs to, and the class of products and services it is involved in, is an important real-world task with many applications in economic, financial, advertising, marketing and other business firms. For instance, such categorization is commonly used for prospective client-lead generation by sales and marketing teams, for  trading strategies like pairs-trading which rely upon highly correlated stocks within the same industry and dealing with similar products, recommending companies to investors and vice-versa based on historical industry and product patterns. Moreover, the knowledge about company's industry, products and services helps in performing industry analysis about the company like general industry economics, industry participants, competition, distribution and buying patterns, etc. 

However, correctly classifying which industries each company belongs to and categorizing its products/services is typically very expensive as it involves manual curation of data about the companies, analyzing all the sectors they might be involved in by going through details of the various products they have built, etc. This quickly becomes a tedious task involving significant amounts of manual labour. Further, it brings about human errors and the results are often dependent on the individual experience and expertise of humans involved in this process. Additionally, with the interdisciplinary nature of the new products and services that companies are creating, a majority of the companies belong to more than one industry, for instance, a company building an educational platform on a mobile app belongs to three industries: education, software and mobile applications, and correctly classifying each company to multiple classes manually is extremely tedious and prone to errors. Furthermore, the lack of a common taxonomy of industry sections and classes of products and services that is useful and agreed upon across different institutions and firms further exacerbates the problem, and necessities such internal development activities at each individual firm. It may be noted that we are not interested in classifying only the available public traded companies, but rather all companies, i.e. public or private, small or large, etc. 

\begin{figure}[b]
\vskip -0.2in
\begin{center}
\centerline{\includegraphics[width=\columnwidth]{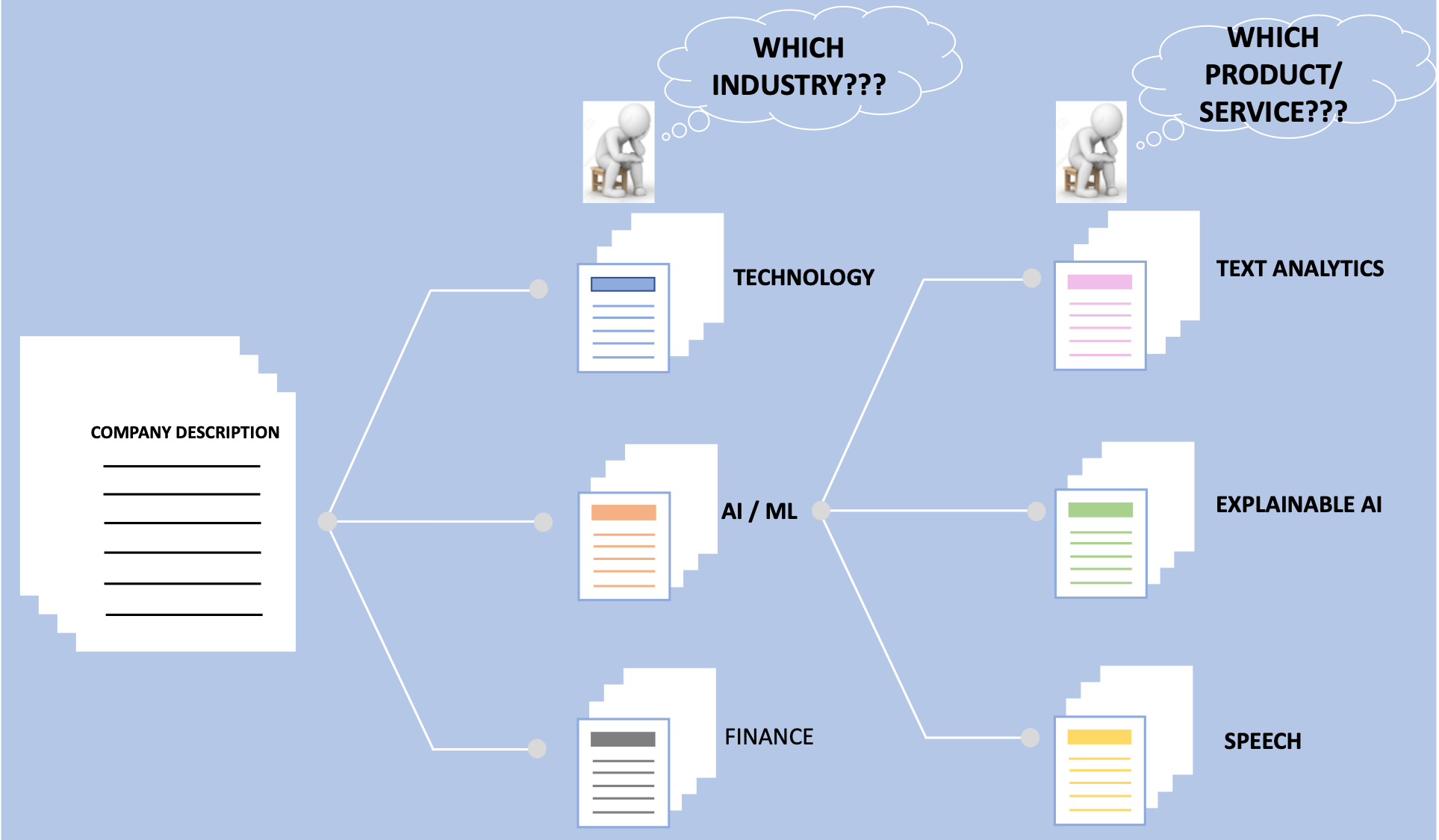}}
\caption{Problem Statement}
\label{fig:Intro}
\end{center}
\vskip -0.2in
\end{figure}

To overcome the above limitations and the need for a human expert with years of experience, we propose to build a system for classifying the company's industries and categorizing its products and services in an automated manner using a hybrid deep-learning based hierarchical classifier. However, most current classification techniques either involve building a supervised multi-class multi-label classifier that require thousands of labeled dataset or unsupervised hierarchical clustering approach which are not easily scalable and have low precision. One of the standard taxonomy datasets for industry code classification is the North American Industry Classification System (NAICS) \cite{NAICScode} codes but due to its lack of adaptation to changing economy, NAICS codes have not gained real-life adoption by most financial organizations. Hence, in case of financial applications, unavailability of labeled dataset and direct financial consequences makes it a hard problem to solve and poses a serious challenge to adoption within the financial domain. In this work, we propose a novel algorithm that leverages recent advances in representation learning to build a hierarchical multi-class classifier along with a targeted multi-label classifier which predicts which industries each company belongs to and which products and services the company builds within each industry by only using company's description as input.

Formally, Fig \ref{fig:Intro} above summarizes our problem. \textit{Given a limited sample-set of companies A, B...N, together with company descriptions, can we correctly predict which custom-defined industry codes each company belongs to and what products and services within each industry the company is involved in?}

Our proposed approach consists of first constructing a labeled dataset for training a multi-class industry classifier by leveraging human subject matter experts (SMEs) and unsupervised techniques. For this work, a pre-defined set of custom named industries and products and services within each industry were provided by the SMEs \footnote{To protect proprietary data, the names of pre-defined set of custom industries and products/services have not been disclosed.}. The SMEs also provided brief one-to-two line description for each of the industry as well as the product and services taxonomy. We then build a high dimensional vector representation for each company using its description as available input data and predict top 3 industries in which the company might belong to. Finally, we construct a targeted multi-class products and services classifier by first building a high dimensional vector representation for each company as well as predefined products and services taxonomy within the predicted industries using their descriptions as inputs and generate similarity scores between them which are then used to identify top 2 products and services that the company might be involved in.

The rest of the paper is outlined as follows. Section 2 describes related work and highlights the unique challenges that come about in building an industry and product/service code classifier, section 3 describes the details of how we constructed the dataset and section 4 provides details of our proposed hierarchical architecture and implementation in more detail. In section 5 we review the results of our approach and also demonstrate how the human targeted SME verification helped in the adoption of our model in the real-world of financial domain. Finally, section 6 concludes our work and lays out foundation for future work in this area.

\section{Related Work}\label{relwork}

There has been numerous works that focus on solving hierarchical text classification which try to leverage recent advances in deep learning and solve the task using supervised techniques such as \cite{yao2019clinical}, \cite{song2020examining}, \cite{luo2021efficient}, and \cite{lavanya2021deep} has done a comprehensive survey on these techniques. While many of these techniques have achieved remarkable results, they require huge amounts of labeled hierarchical data which is very hard to obtain in the financial domain. As mentioned in Section \ref{intro}, one of the available taxonomy datasets for industry code classification is the NAICS \cite{NAICScode} codes, a six-digit code, containing 1057 codes which comprises of 20 sectors, 99 sub-sectors, and 311 industry groups. However, this dataset is very focused on North America economy and has not been adaptive enough to capture the changing economy and hence has not been adopted by most financial organizations. This makes the problem very hard to solve as every financial organization defined their own custom industry and products and services codes.

Some groups have tried to solve the text classification problem by using unsupervised techniques like hierarchical clustering and topic modeling such as \cite{aljedani2021hmatc}, \cite{zhang2021match}, \cite{rafea2021classification}. However, these models are hard to get adopted in real-world financial domain as they do not perform well on large scale data and the low precision results can have direct financial consequences as explained in Section \ref{intro}.

Additionally, there also has been some work in the literature on automated industry classification \cite{10.1007/11552413_14}, \cite{doi:10.1080/08963568.2015.1110229}, \cite{https://doi.org/10.48550/arxiv.2004.01496}, almost all these approaches rely on self- reporting, manual entry, and possibly naive algorithms. \cite{wood2017automated} and \cite{oehlert2022naics} explored the possibility of using NAICS code for building a supervised learning algorithm and leverage advancements in deep learning. \cite{papagiannidis2018identifying} tried to use clustering algorithms to capture similarities across companies and identify industry codes. These methodologies are not scalable and are difficult to be adopted and deployed in financial domain since the companies these days belong to multiple industries and the products and services are interdisciplinary in nature. Moreover, there has been no literature on hierarchical classification of the companies into identifying what products and services the companies are involved in. In this work, we have tried to leverage advances in deep learning and unsupervised learning to build a hierarchical multi-class multi-label classifier which uses the company's descriptions as input and predicts top 3 industries that the company might belong to and top 2 product/services within each industry that the company might be involved in.

\section{Data}

This section describes the details for our dataset. Our model has been developed and tested using licensed data from Pitchbook \cite{Pitchbook.2021}. The dataset contains details on companies such as a brief description on what the company does as well as pitchbook-defined industry verticals. Moreover, we were provided with $\sim$40 custom defined industry codes and $\sim$400 products and services codes by the SMEs  into which the companies had to be classified in. The following sections detail how we constructed the dataset to build an industry code classifier as well as the products and services classifier.

\subsection{Dataset Construction for Industry Code Classification}\label{iccdata}

To classify the companies with high accuracy and precision into which industries they belong to, we first construct a labeled dataset with input as company descriptions and output as the industry codes. In order to construct this dataset, we leveraged the pitchbook industry codes. The pitchbook industry codes are defined in three levels: (a)Primary Industry Sector: A broad category that contains industry groups and codes, (b)Primary Industry Group: A sub-category that provides more specific classification, (c)Primary Industry Code: The primary industry the company operates in.

In order to obtain the mapping between pitchbook descriptions and SME's custom defined industry codes, we leveraged our SMEs to create a mapping between the pitchbook's three level codes and SME's industry codes. Through this mapping we were able to generate a labeled dataset containing pitchbook descriptions as inputs and SME's industry codes as output. However, through this mapping we were only able to generate the labeled dataset for 75\% of the industry codes.

\begin{figure}[t]
\begin{center}
\centerline{\includegraphics[width=\columnwidth]{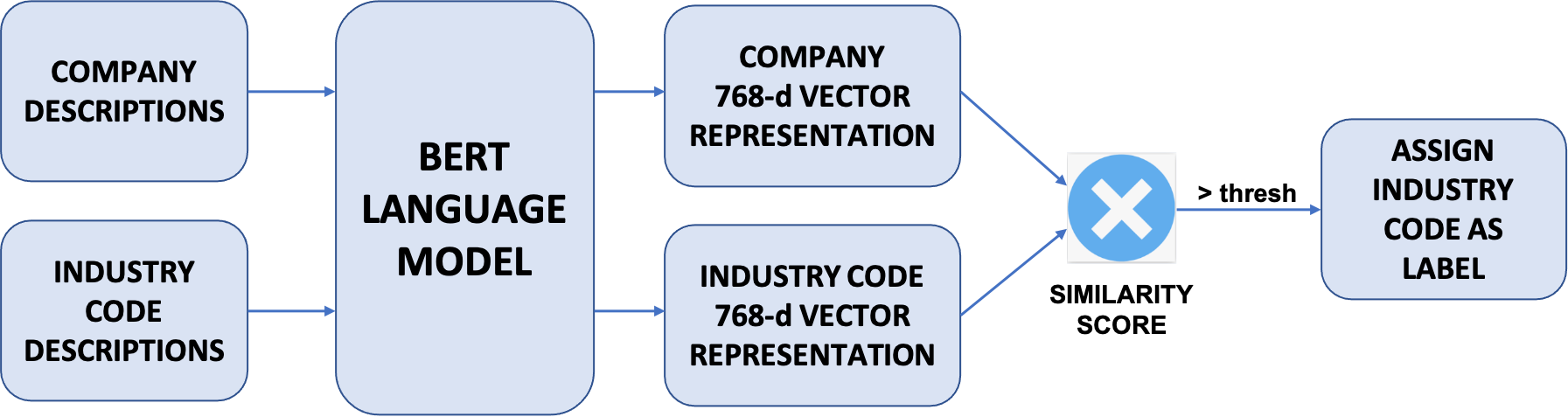}}
\caption{Construction of Labeled Data for the remaining 25\% of the industry codes for Industry Code Classification}
\label{fig:data}
\end{center}
\vskip -0.3in
\end{figure}


For the remaining 25\%, we used an unsupervised approach to obtain the labeled dataset. Fig \ref{fig:data} outlines the methodology used to construct the dataset for the remaining industry codes. Since the industry codes in itself do not contain much contextual meaning, the SME's provided us with brief one-to-two line descriptions of all the custom defined industry codes. In order to enable us to easily perform comparisons and estimate how close/far the companies and the industry codes are, we then used a pre-trained large language model to encode the company descriptions and industry code descriptions into a high dimensional distributed vector representation. The key guiding principle behind the high dimensional vector representation is that similar sentences/words, or phrases within a similar contexts should map close to each other in a high-dimensional space, and sentences/words or phrases that are very different should be far away from each other. We then estimate the similarity scores between the two vector representations using cosine similarity, $cos(comp, ind\_code)$. The industry code which has the highest similarity score and is above a given threshold, $thresh$ (hyperparameter), is then assigned as the label to the corresponding company description. Please note that these comparisons are performed only for the remaining 25\% of the industry codes.
\vspace*{-2mm}
\begin{equation}
    cos(comp, ind\_code) = \frac{comp\_vec.ind\_vec}{||comp\_vec||.||ind\_vec||}
\label{cosine-similarity}
\end{equation}


\subsection{Dataset Construction for Products and Services Code Classification}\label{psdata}

Since we use an unsupervised approach for this classification, we only used the company descriptions from pitchbook dataset. As the product and service codes in itself do not contain much contextual meaning, the SME's provided us with one-to-two line descriptions of all the custom defined product and service codes.

\section{Proposed Approach}

This section describes our algorithm used to classify the companies into the industry codes to which it belongs, in Subsection \ref{icc}, and subsequently how it predicts the product and service codes in which the companies are involved in, in Subsection \ref{pscc}. 

\begin{figure}[t]
\begin{center}
\centerline{\includegraphics[width=\columnwidth]{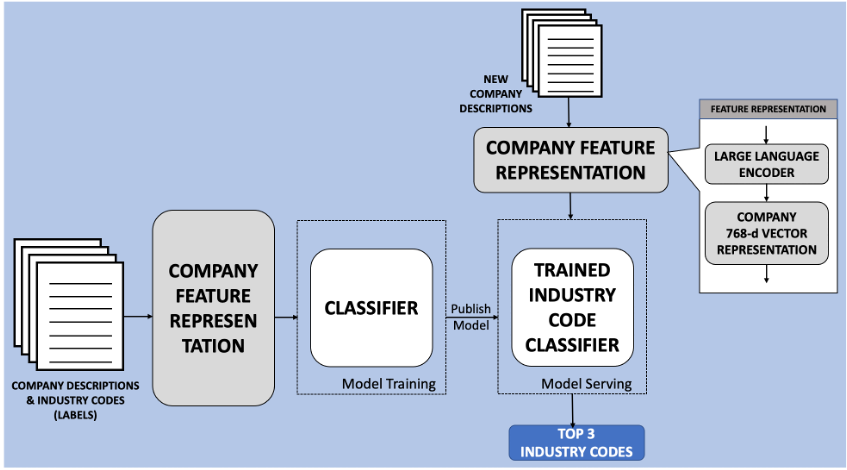}}
\caption{Industry Codes Classification Methodology}
\label{fig:Level1Model}
\end{center}
\vskip -0.2in
\end{figure}

\subsection{Industry Code Classifier} \label{icc}

Figure \ref{fig:Level1Model} describes our proposed approach to solve the industry code classification problem i.e. \it 'What are the industry codes to which Company $i$ belongs to?' \rm. Our proposed approach consists of two sequential steps. In the first step, we pass the company descriptions through a pre-trained large language model, specifically Roberta-base model \cite{https://doi.org/10.48550/arxiv.1907.11692}, and obtain high dimensional distributed vector representations for each company. In the second step, we use these representations as input features to train a multilayer perceptron with a classification layer using the labeled dataset constructed in Section \ref{iccdata}. This trained classification model is then used to predict industry codes for new companies. Moreover, since the Roberta-base model is pre-trained over generic Wikipedia data, this makes our trained classifier generalizable to predict industry codes for company descriptions obtained from any data source.

\subsection{Products and Services Code Classifier} \label{pscc}

\begin{figure}[b]
\begin{center}
\centerline{\includegraphics[width=\columnwidth]{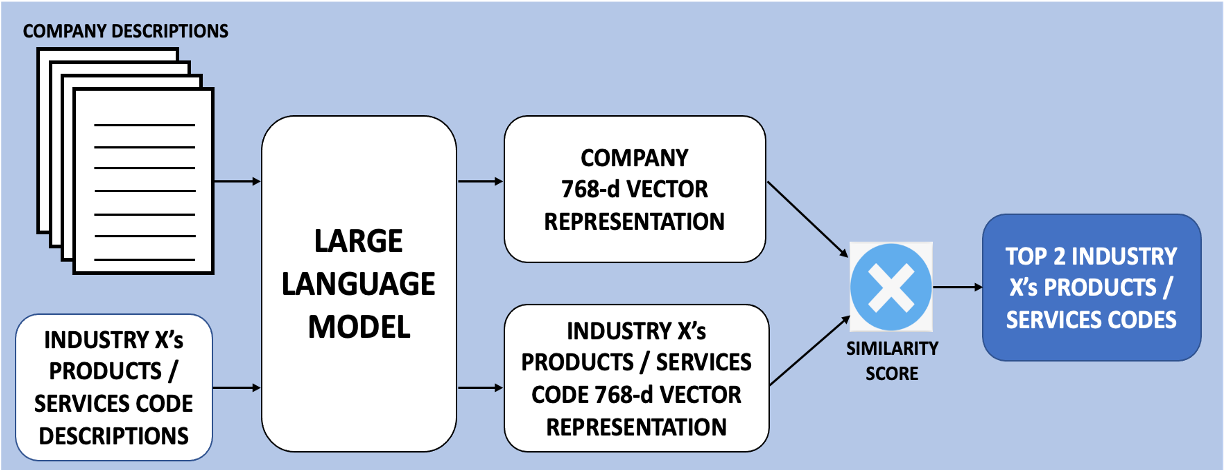}}
\caption{Product/Service Codes Classification Methodology}
\label{fig:Level2Model}
\end{center}
\vskip -0.3in
\end{figure}

Figure \ref{fig:Level2Model} describes our proposed approach to solve the product and service code classification problem i.e. \it 'What are the product and services codes within a particular industry that the Company $i$ is involved in?' \rm. This is the second targeted classification layer wherein we try to predict which product and service codes the company is involved in within each predicted industry code ($\sim$8-15 product/services codes within each industry) in an unsupervised manner, obtained from Subsection \ref{icc}. In this approach, we obtained the products and services code descriptions within each industry code from the SMEs, as described in Section \ref{psdata}. In order to enable us to easily perform comparisons and estimate how close/far the companies and the product and service codes are, we then used a pre-trained large language model, specifically Roberta-base model, to encode the company descriptions and product and service code descriptions into a high dimensional distributed vector representation. We then estimate the similarity scores between the two vector representations using cosine similarity, Equation \ref{cosine-similarity}. The product and service codes which have the top 2 similarity scores are then assigned as the predicted product and service codes to the corresponding company. 

\noindent The \it Algorithm \ref{alg1} \rm describes our overall industry and product/service codes prediction algorithm in a step-by-step manner. 

\vskip -0.1in
\begin{algorithm}
 \caption{Industry \& Product/Service Codes Prediction Algo}
 \label{alg1}
 \begin{algorithmic}[1]
  \STATE \textbf{Input 1:} Companies, say $comp$ where $comp\in [1, 2, 3, \hdots, m]$, its descriptions, say $desc$ where $desc\in [1, 2, 3, \hdots, m]$ and its labels, say $l$ where $l\in [1, 2, \hdots, m]$, where $m$ is the number of companies in the dataset 
  \STATE \textbf{Input 2:} Products and services taxonomy, say $ps\_code$ where $ps\_code\in[A,B, \hdots, N]$ and its descriptions, say $ps\_code\_desc$ where $ps\_code\_desc\in[A,B, \hdots, N]$, where $N$ is number of products and services codes belonging to each industry code ($N$ varies for each industry code)
 \STATE \textbf{(1) Industry Codes Classifier:}
 \STATE Split the dataset into training set, say $comp\_tr, desc\_tr, l\_tr \in [1, 2, 3, \hdots, k]$, and test set, say $comp\_ts, desc\_ts, l\_ts \in [1, 2, 3, \hdots, k]$
 \FOR {$comp\_tr, desc\_tr=[1, 2, \hdots, k]$}
 \STATE Obtain $tr\_v=$ vector rep. for $desc\_tr$
 \ENDFOR
 \STATE Train classification model, say $ind\_code\_mo$, using $tr\_v$ as inputs and $l\_tr$ as actual labels
 \FOR {$comp\_ts, desc\_ts=[1, 2, \hdots, k]$}
 \STATE Obtain $ts\_v=$ vector rep. for $desc\_ts$
 \STATE Use $ind\_code\_mo$ and predict top 3 industry codes for $comp\_ts$, say $comp\_InC\in [X, Y, Z]$
 \ENDFOR
 \STATE \textbf{(2) Products and Services Codes Classifier:}
 \FOR {$comp\_ts, desc\_ts=[1, 2, \hdots, k]$}
 \FOR {$comp\_InC=[X, Y, Z]$}
 \FOR {$ps\_code=[A,B, \hdots, N]$}
    \STATE Obtain $ps\_v=$ vector rep. for $ps\_code$ description
    \STATE Obtain $ts\_v=$ vector rep. for $desc\_ts$
    \STATE Calculate cosine similarity,\\ $sim(comp, ps\_code) = \frac{ts\_v.ps\_v}{||ts\_v||.||ps\_v||}$
\ENDFOR
\STATE Obtain top 2 $prod\_codes$ for which the $sim(comp, ps\_code)$ are the highest
 \ENDFOR
 \ENDFOR
 \end{algorithmic}
  \vskip -0.03in
 \end{algorithm}

\section{Experiments and Results}

\begin{figure}[t]
\begin{center}
\centerline{\includegraphics[width=\columnwidth]{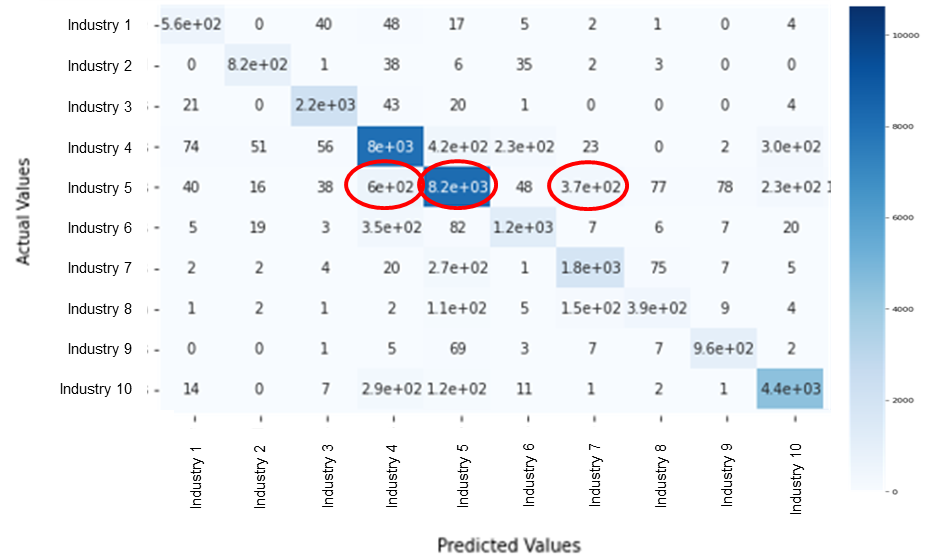}}
\caption{Confusion Matrix for a sample of 10 industry codes}
\label{fig:ConfMatr}
\end{center}
\vskip -0.2in
\end{figure}

In order to first evaluate the performance of our industry code classifier, we first split the labeled dataset constructed in Section \ref{iccdata} into 80\% training and 20\% test set. Moreover, as discussed in Sections \ref{intro} and \ref{relwork}, since the companies are interdisciplinary in nature we generated the confusion matrix and observed that majority of the companies cover the span of at least 3 industry codes. Fig \ref{fig:ConfMatr} shows the confusion matrix for a sample of 10 industry codes\footnote{Due to proprietary reasons, the industry codes have been masked.} with red circles depicting the span size for $Industry$ $Code$ $5$ . Finally, Table \ref{tab: res} shows top-3-accuracy score of the industry code classifier on this labeled test set.

Further, in order to test the accuracy of product and services classifier as well as for real-life adoption, we were provided a small set of 350 companies which were manually labeled by Subject Matter Experts (SMEs) containing the ground truth for both industry and product and service. Table \ref{tab: res} shows top-3-accuracy score of industry classifier and top-2-accuracy score of product and services classifier on this SME-labeled dataset. 

\begin{table}[h]
\centering
\begin{adjustbox}{width=\columnwidth,center}
\begin{tabular}{c|c|c|c}
\hline
                                                                & \begin{tabular}[c]{@{}c@{}}Industry Code \\ Classifier\end{tabular} & \begin{tabular}[c]{@{}c@{}}Products and \\ Services Classifier\end{tabular} & \# of Samples \\ \hline
\begin{tabular}[c]{@{}c@{}}Constructed \\ Test Set\end{tabular} & 92.38\%                                                             & -                                                                           & 20,000        \\
\begin{tabular}[c]{@{}c@{}}SME-Labeled \\ Dataset\end{tabular}  & 94.87\%                                                             & 96.85\%                                                                     & 350           \\ \hline
\end{tabular}
\end{adjustbox}
\caption{Results of Industry Code Classifier and Product and Service Code Classifier}
\label{tab: res}
\vskip -0.4in
\end{table}

\section{Conclusion and Future Work}

Our work proposes a novel hierarchical classification algorithm that leverages recent advances in representation learning to predict high precision custom-defined industry codes as well as products and services codes for a particular company by only using
company’s description as input. We also highlight how a high quality labeled industry code dataset can be constructed through mapping and unsupervised techniques, and demonstrate how a targeted unsupervised multi-class products and services classifier can yield high precision predictions resulting in real-life adoption and deployment. 

The work opens numerous avenues to further build upon. For instance, we could further enhance the algorithm to automatically generate custom industry codes and product and services codes based on company descriptions and then using public information extraction techniques to extract code definitions hence eliminating the need for predefining these codes by a subject matter expert. 

\bibliographystyle{ACM-Reference-Format}
\bibliography{InProc}

\appendix

\end{document}